%% file: main.tex
\documentclass[fleqn,10pt]{wlscirep}

\usepackage{chngcntr}
\usepackage[utf8]{inputenc}
\usepackage[T1]{fontenc}
\usepackage{setspace}
\usepackage{nameref}
\usepackage{url}

\usepackage[colorlinks=true]{hyperref}

\begin{document}
\title{Measuring Technological Complexity Using Japanese Patents}
\author[1]{Rintaro Karashima}
\author[1, 2, *]{Hiroyasu Inoue}
\affil[1]{University of Hyogo, Graduate School of Information Science, Kobe, Hyogo 6500047, Japan}
\affil[2]{RIKEN, Center for Computational Science, Kobe, Hyogo 6500047, Japan}

\affil[*]{inoue@gsis.u-hyogo.ac.jp}


\begin{abstract}
As international competition intensifies in technologies, nations need to identify key technologies to foster innovation. 
However, the identification is challenging due to the independent and inherently complex nature of technologies. 
Traditionally, analyses of technological portfolios have been limited to simple evaluations, indicating merely whether a technology is specialized. 
We propose evaluating TCI at the corporate level, which provides finer granularity and more detailed insights than conventional regional evaluations by using Japanese patent data spanning fiscal years 1981 to 2010. 
Specifically, we analyze a bipartite network composed of 1,939 corporations connected to technological fields categorized into either 35 or 124 classifications. 
Our findings quantitatively characterize the ubiquity and sophistication of each technological field, reveal detailed technological trends reflecting broader societal contexts, and demonstrate methodological stability even when employing finer technological classifications. 
Additionally, our corporate-level analysis allows consistent comparisons across different regions and technological fields, clarifying regional advantages in specific technologies.
The corporate level analysis also reveals a new possibility for formulating an innovation strategy.

\end{abstract}

\flushbottom
\maketitle

\thispagestyle{empty}

\input{Sections/introduction.tex}

\input{Sections/results.tex}

\input{Sections/discussion.tex}

\input{Sections/methods.tex}


\input{Sections/data_availability.tex}
\input{Sections/code_availability.tex}

\bibliography{ref}


\input{Sections/author_contributions.tex}

\section*{Competing interests}
The authors declare no competing interests.

\input{Sections/acknowledgements.tex}

\newpage
\input{Sections/supplementary_information.tex}

\end{document}

%% file: Sections/introduction.tex
\section*{Introduction}
\phantomsection
\addcontentsline{toc}{section}{Introduction}
\label{section:Introduction}

In the twenty-first century, intensifying international competition has made technological capability a central element of national competitiveness. 
Nations seek to foster environments that support the development of sophisticated technologies, as these capabilities are fundamental to sustainable economic growth. \cite{EC_SSP_nodate,Teixeira2022,NISTEP_nodate}
The ecosystems that produce such capabilities are, however, inherently complex, involving interactions among countries, corporations, and technologies \cite{Weaver1948science}. 
This complexity makes it difficult to accurately identify the key technologies and actors within the system. 
Understanding this structure is crucial for developing effective science and technology strategies to strengthen a nation's competitive advantage.

Historically, the assessment of technological capabilities has predominantly relied on quantitative metrics derived from bibliographic data of patents and publications, ranging from simple patent counts to more complex citation-based indicators, including Technology Relevance (TR) \cite{jaffe2019patent,aristodemou2018citations}. While such indicators offer certain insights by considering citation practices, patent age, and technological fields, they suffer from significant shortcomings. Their inherent dependence on the long-term accumulation of citations makes them susceptible to biases from citation behaviors, patent examiners, and strategic patenting. Crucially, these metrics fail to capture the core aspect of technological complexity, namely the intricate interplay among corporations and technologies. Consequently, approaches that rely solely on these data provide a limited perspective and inadequately reflect the qualitative dimensions of technological capability.

The Hidalgo-Hausmann algorithm\cite{hidalgo2009eci} provides a powerful solution by considering both the diversity of a region and the ubiquity of its products or technologies. 
This dual perspective allows the methodology to quantify their complexity, revealing detailed economic structures not captured by traditional metrics. 
The utility of this foundational framework has led to its extension into several distinct streams of research\cite{Hidalgo2021,Hidalgo2023,Hausmann2024,Balland2022,Chakraborty2020}. 
One major development is the introduction of complementary metrics such as fitness-based approaches \cite{Tacchella2012,Wu2016,Albeaik2017}. 
Another significant stream, which is the focus of this study, involves the adaptation of the algorithm to patent data, pioneered by Balland and Rigby to measure technological complexity \cite{balland2017tci}. 
Building on this work, recent studies have further extended the patent-based analysis not only to Europe\cite{Pinheiro2022,ballandandboschma2021}, but also to Asian economies \cite{Jun2023,Abay2024}.
Despite this wide range of applications, the focus has predominantly remained at the regional level \cite{Hidalgo2021}. 
Therefore, shifting the analytical focus to the corporate level, holds the potential to unlock finer-grained comparisons of technological capabilities and regional specializations, revealing dynamics that a regional-level view cannot capture.

Our study applies the HH algorithm to an extensive set of Japanese patent records to quantify technological complexity at the corporation level. 
By constructing bipartite networks linking corporations to technology classes, we develop a detailed measure of technological sophistication. 
Our analysis investigates the relationships between technological complexity, ubiquity, and the average corporate diversity of patent technologies, providing new insights into the nature of advanced capabilities. 
The model developed at the corporation level provides a more robust and consistent assessment of technology than regional-level analyses, particularly for finely classified technological fields. 
Importantly, this study shows that a corporation-level approach enables a direct comparison of technological characteristics across different regions. 
By focusing on corporations within specific regions, we can identify distinct industrial specializations and reveal which technologies are complex within each region’s unique context. The analyses from this study provide insights for creating tailored industrial policies and more accurately identifying the key technologies vital for future growth.

The rest of this paper is organized as follows. 
In \nameref{section:Results} section, we present the ubiquity and complexity of technological fields presented by the patent data, followed by a detailed analysis of technological trends.
In the \nameref{section:Discussion} section, we provide a critical analysis of our findings and discuss the implications of our results.
Finally, in the \nameref{section:Methods} section, we detail the data processing and methods used in our analysis.

%% file: Sections/results.tex
\makeatletter
\def\@currentlabelname{Results}%
\makeatother
\label{section:Results}

\section*{Results}
\phantomsection
\addcontentsline{toc}{section}{Results}

\input{Sections/Results/characteristics.tex}

\input{Sections/Results/consistency.tex}

%% file: Sections/Results/characteristics.tex
\subsection*{Characteristics of Technologies} 
\phantomsection
\addcontentsline{toc}{subsection}{Characteristics of Technologies}
\label{subsection:characteristics}

In this study, we use the HH algorithm to quantify the structural characteristics of technologies based on corporate patent publications from 1,939 corporations filed between fiscal years 1981 and 2010 (see \nameref{section:Methods} section).
First, we define ubiquity (or degree centrality) $K_{T,0}$ as the number of corporations connected to a technology which lower value indicates that the technology is more rare. 
Similarly, the average diversity $K_{T,1}$, which measures the variety of corporations connected to a technology, implies that a higher value reflects diversified publication, whereas a lower value suggests specialization. 
Generally, the average of each indicator is used as a threshold to distinguish high from low values \cite{hidalgo2009eci,balland2017tci, Balland2018}.

Another key indicator is TCI, which represents the sophistication of a technology by reflecting how advanced the knowledge required for its production. A threshold of zero divides TCI into high and low categories: a high TCI implies that a technology demands a complex combination of unique inputs and advanced technical expertise which indicates only a few advanced corporations can publish it competitively, whereas a low TCI indicates reliance on widely available capabilities, allowing many corporations to participate.

\begin{figure}[ht]
  \centering
  \includegraphics[scale=0.95]{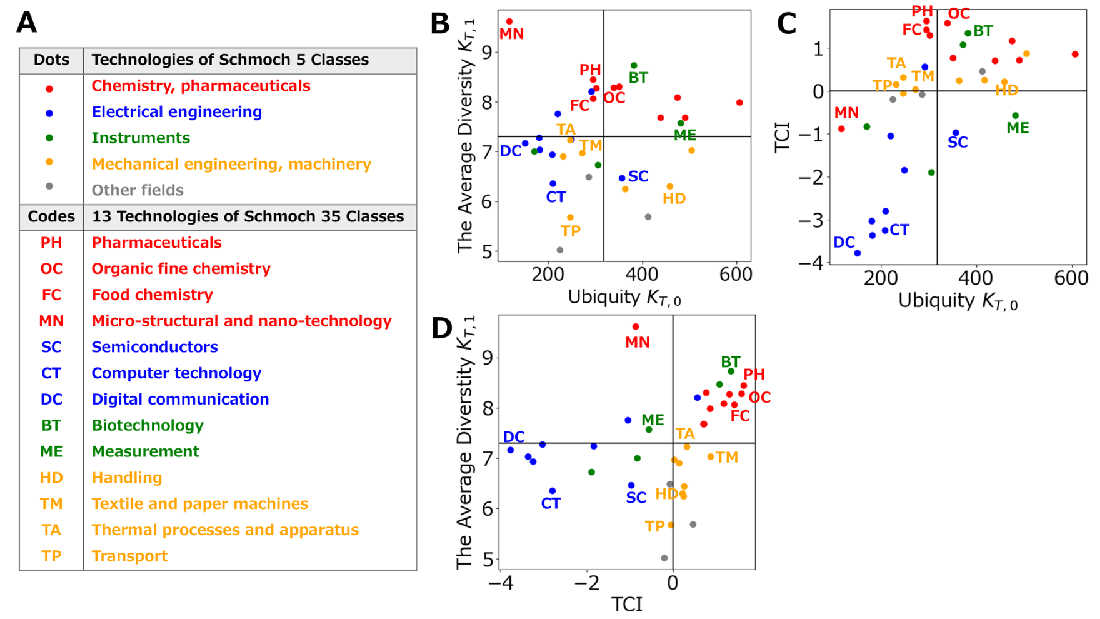}
  \caption{
  (A) The legend of the five classifications defined by Schmoch \cite{Schmoch2008}, along with an explanation of the technology codes used in Panels (B), (C), and (D).
  (B) The Ubiquity \(K_{T,0}\) versus average diversity \(K_{T,1}\).
  (C) The Ubiquity \(K_{T,0}\) versus TCI.
  (D) The TCI versus average diversity \(K_{T,1}\).
  The black lines in Panels (B), (C), and (D) indicate the mean values of \(K_{T,0}\) and \(K_{T,1}\), and \(\text{TCI} = 0\) is also shown with a black line. Each dot is color‐coded according to (A).
  }
  \label{fig:scatter}
\end{figure}

To systematically evaluate technological characteristics, we map the relationships among ubiquity $K_{T,0}$, average diversity $K_{T,1}$, and TCI across patent technologies (Fig.~\ref{fig:scatter}).
Technologies are classified according to the Schmoch system \cite{Schmoch2008}, as illustrated in Panel A of Fig.~\ref{fig:scatter}. Values for ubiquity and average diversity are normalized by their respective means, while TCI is segmented by the zero threshold.

We find a negligible correlation between ubiquity $K_{T,0}$ and average diversity $K_{T,1}$ (Pearson correlation coefficient $r=0.064$), indicating that widely disseminated technologies are not necessarily associated with diversified corporate portfolios (Panel B in Fig.~\ref{fig:scatter}). Technologies within Electrical Engineering, for example, typically show low ubiquity and low average diversity, indicating concentrated innovation within narrowly defined technological domains.

Conversely, ubiquity $K_{T,0}$ and TCI exhibit a moderate positive correlation (Pearson correlation coefficient $r=0.594$), suggesting that technologies associated with fewer corporations often require complex capabilities (Panel C in Fig.\ref{fig:scatter}). 
In this context, Electrical Engineering exhibits low ubiquity, low average diversity, and low TCI values, indicating that corporations in this field tend to concentrate their innovation efforts within narrowly defined domains.

Detailed analysis using International Patent Classification (IPC) classes further supports these observations (Fig.~\ref{fig:persector}). 
Within the five major sectors defined by Schmoch \cite{Schmoch2008}, the relationship between ubiquity and average diversity is predominantly weak or moderate. Specifically, Electrical Engineering displays a strong positive correlation between ubiquity and TCI (Pearson correlation coefficient $r=0.738$), highlighting the coexistence of corporations specializing in broadly applicable technologies and those focusing on niche, specialized innovations.

Electrical engineering technologies are predominantly located in the region of low TCI and low average diversity (Panel B in Fig.~\ref{fig:scatter}). In contrast, Chemistry and Pharmaceuticals technologies are clustered in the area of high TCI and high average diversity, exhibiting a positive correlation between these two metrics (Panel D in Fig.~\ref{fig:scatter}; Pearson correlation coefficient $r$=0.316). Furthermore, corporations in the Chemistry and Pharmaceuticals domains are connected not only to fields with low ubiquity but also to those that are more common such as Biotechnology, Analysis of Biological Materials, and Measurement (Panel C in Fig.~\ref{fig:scatter}).
Thus, these results demonstrate a fundamental difference between the two technological domains. The Chemistry and Pharmaceuticals sectors are characterized by rarity and high sophistication and are typically developed by more diversified corporations. A detailed examination reveals a slight negative correlation between ubiquity and average diversity (Fig.~\ref{fig:persector}), suggesting that in sectors with high technological complexity, corporations with more comprehensive technological portfolios tend to produce rare technologies.
Furthermore, the IPC classes corresponding to Biotechnology and Measurement show high TCI values in fields related to the production of sugar and yeast (C13, A21, C12), drug discovery (A61, C07), and the preservation of these technologies (C07, A23). These findings are consistent with the socio-technical context of the period, particularly reflecting the emergence of drug discovery technologies in the 1980s and 1990s \cite{Sakakibara2014,Nakamura2022}. Moreover, the significant roles of food chemistry and biotechnology during the 1990s and 2000s underscore their contribution to the development of additives that enabled safer food preservation and processing not only in Japan but globally \cite{Bhatia2018,Murakami2024}.

%% file: Sections/Results/consistency.tex
\subsection*{Consistency over Time}
\phantomsection
\addcontentsline{toc}{subsection}{Finer Technological Trends and Consistency}
\label{subsection:consistency}
The preceding section focuses on the corporate TCI over the full period from fiscal year 1981 to 2010 and provides a baseline view of technological characteristics in Japan. 
In this section, we explore whether more granular insights into regional technological capabilities can be obtained by comparing the regional- and corporate-level TCIs.
Specifically, through comparing results derived from the Schmoch classes and the three-digit IPC classes, we examine how classification granularity affects the consistency of technological complexity evaluations across regions and corporations.

The two classification systems share the same technological scope but differ in their level of aggregation. 
The Schmoch system includes 35 broad technology classes, whereas the three digit IPC used in this study contains 124 finer classes. 
Previous studies on regional technological complexity have reported that using finer classifications tends to make the bipartite network sparse and to reduce the consistency of the calculated indicators\cite{morrison2017economic,Tacchella2012}. 
Such inconsistency partly reflects the limited number of regional nodes relative to the number of technology classes. 
In regional analyses, several dozen to a few hundred regions such as countries or EU cities\cite{Hidalgo2021,nepelski2020technological} are connected to several hundred or thousand technology classes, which limits the variation in ubiquity across technologies and produces a biased distribution of TCI. 
By contrast, corporate analyses involve a much larger number of nodes, often exceeding several thousand corporation, and therefore capture a broader range of ubiquity through the iterative calculation.

\begin{figure}[ht]
  \centering
  \includegraphics[scale=1.5]{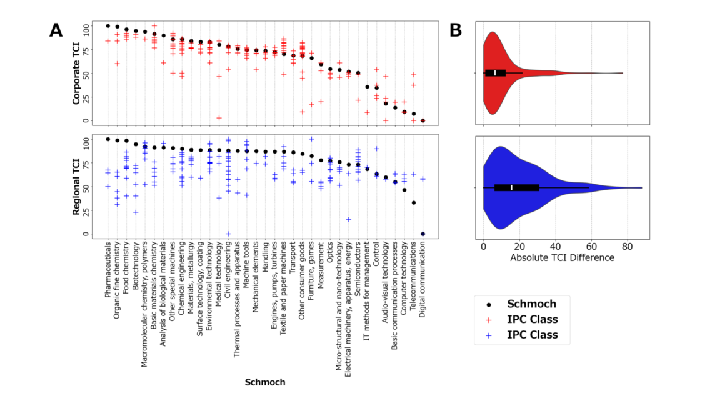}
  \caption{
    (A) The TCI scores of 124 IPC the first three-digit classes and 35 Schmoch classes for both prefectures and corporations. The scores have been scaled to values ranging from 0 to 100. Each IPC is plotted against the corresponding Schmoch axis, and in
cases of multiple correspondences (i.e., Schmoch class “Chemical Engineering” maps to IPC classes B01–B08, C14, etc.), IPC classes are plotted individually. 
    (B) The distribution of the absolute differences between TCI of IPC and of Schmoch classes.
  }
  \label{fig:detailtci}
\end{figure}

These tendencies are reflected in the observed distributions of TCI.
In the regional analysis based on the Schmoch system, 26 out of 35 classes fall within the range of 75 to 100 (Fig.~\ref{fig:detailtci}), indicating that most technologies exhibit similar levels of complexity.
This limited dispersion is consistent with the small number of regional nodes, which comprises only 47 prefectures in Japan.
In contrast, the corporate analysis involves 1,939 corporations, resulting in a denser bipartite network that captures greater heterogeneity in technological capabilities.
Consequently, only 15 classes lie within the same range in the corporate case, while the remaining classes show a wider spread of TCI values across technologies.
This broader dispersion suggests that the corporate analysis captures the variation in ubiquity more effectively, revealing distinct differences in how frequently each technology appears across corporations.

The difference between the approaches becomes even clearer when using a finer classification.
When the classification becomes more detailed and the number of technology nodes increases, the regional approach cannot define ubiquity sufficiently.
In the case of Civil engineering, the variance of TCI values for IPC classes corresponding to the same Schmoch class is larger than that for the corporate approach.
The larger difference between the TCI values of the coarse classification and those of the finer classification indicates a greater inconsistency in the evaluation for the regional approach.
A Mann-Whitney U test shows that the distribution of residuals between the TCI values of the Schmoch and IPC classes is significantly different (U = 7495.5, $p$ value = 6.55e-15, $\gamma$ value = 0.60), and that the residuals for the corporate analysis are smaller than those for the regional analysis (Panel B in Fig.~\ref{fig:detailtci}).
Smaller residuals imply that the corporate TCI is less affected by the choice of classification level and provides a more consistent measure of technological complexity.
This result suggests that the corporate network captures the co-occurrence patterns of technologies more precisely because of the larger number of corporation nodes and the finer granularity of technological connections.
Hence, the corporate analysis offers a more stable representation of how diverse and rare capabilities are combined in actual technological activities.
These consistent and detailed trends are observed not only over the 30-year period but also in shorter five-year intervals (SI Fig.~\ref{fig:bump}).
For example, in the IPC classes related to Food Chemistry, a persistently high TCI is observed for sugars (C13) and baking technologies (A21) throughout the 2000s.
By contrast, foodstuffs (A23) and agriculture-related technologies (A01) show gradual declines in ranking, particularly after the mid-1990s.
This indicates that even within related technological fields, some areas maintain their complexity advantage while others lose ground.
These results confirm that the relative positioning of technologies can be captured in a stable and systematic manner across different time resolutions.

\begin{figure}[ht]
  \centering
  \includegraphics[scale=0.45]{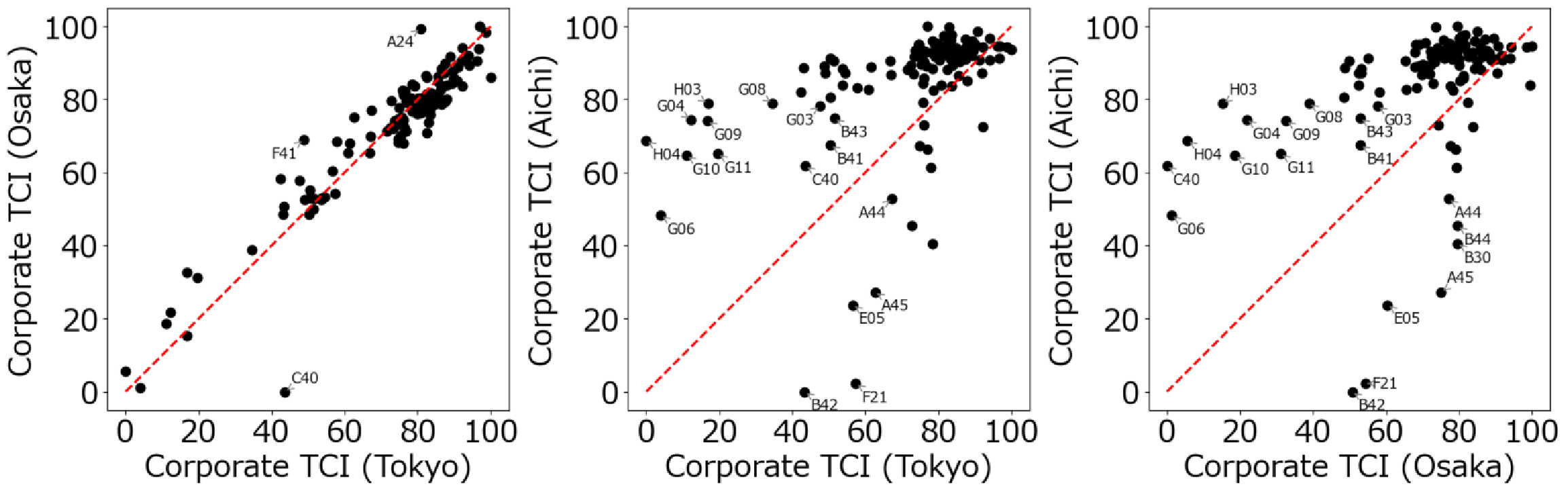}
  \caption{Three scatter plots arranged in a single row compare the Corporate TCI between pairs of prefectures. The left panel compares Osaka with Tokyo, the middle panel compares Aichi with Tokyo, and the right panel compares Aichi with Osaka. Each axis ranges from 0 to 100. Black points represent individual classifications or entities, and the red dashed line indicates the diagonal reference line.}
  \label{fig:prefcompare}
\end{figure}

Corporate analyses are useful not only for examining technological structures at finer classification and temporal scales but also for capturing spatial characteristics in greater detail than regional analyses.
While previous regional studies have mainly compared aggregate indicators such as the average TCI or the presence of links to certain technologies, the corporate analysis enables a direct comparison of each technology's TCI between different regions.
Regional comparisons are informative because similarity or divergence in corporate TCI across prefectures can reveal distinctive industrial bases and specialization patterns.
We compare cross-prefectural alignment by pairing Tokyo, Osaka, and Aichi, and interpret the class-level correspondence from the concentration around the diagonal and notable deviations in the scatter plots (Fig.~\ref{fig:prefcompare}).
Tokyo and Osaka exhibit a close correspondence across many fields, which reflects a strong resemblance in their technological structures.
Tokyo and Aichi show a looser association in which Aichi tends to be stronger in electronics and ICT, for example Computing (G06) and Telecommunications (H04), while Tokyo shows relative strengths in consumer goods and light manufacturing, for example Printing and Bookbinding (B42) and Lighting Technology (F21).
The pair of Osaka and Aichi diverges the most, with Aichi again relatively strong in electronics, for example Computing (G06) and Basic Electronic Circuitry (H03), whereas Osaka appears higher in manufacturing-oriented areas, for example Presses (B30) and Decorative Arts and Processes (B44).
This pattern may highlight that Aichi’s technological profile diverges not only from Tokyo but also from Osaka, reflecting a distinctive regional specialization in electronics and ICT.
The corporate analysis thus makes finer regional technological characteristics visible within each prefecture than the regional-level analysis.

%% file: Sections/discussion.tex
\makeatletter
\def\@currentlabelname{Discussion}%
\makeatother
\label{section:Discussion}

\section*{Discussion}
\phantomsection
\addcontentsline{toc}{section}{Discussion}
This study calculates the Technological Complexity Index (TCI) using Japanese patent data identify sophisticated technologies critical to innovation, and provides three primary contributions.
First, it demonstrates methodological robustness at the corporate level.
Unlike regional approaches, corporate-level TCI retains consitent technological rankings even when classifications become highly detailed.
Our results reveal notable complexity in the Pharmaceuticals and Chemistry sectors, likely reflecting Japan's industrial focus during the pharmaceutical industry's formative period.
This consistency across classification scales represents significant progress compared to previous regional analyses that lost reliability when switching from broader to finer classifications.
Second, by visualizing the temporal evolution of rankings, this study clarifies which technologies appear to have sustained innovative activity during Japan’s prolonged period of modest economic growth. The corporate level lens makes these trajectories easier to read because rankings remain stable enough to track long run patterns while still capturing meaningful changes. This perspective highlights fields that persist at high complexity as well as those that rise or fade, providing a more concrete basis for interpreting how Japan’s innovation system has been supported over time.
Third, the corporate level approach also enables a finer view of regional technological characteristics. By attributing firms to places and then examining their technological portfolios, we can recover local profiles that are obscured by coarse regional aggregates. This yields clearer distinctions among prefectures and urban areas and helps identify locally salient capabilities that may be missed when analysis is performed only at the regional level.

At the same time, there are limits that point to future work. Detecting dynamic shifts requires a balance between longer windows that improve stability and shorter windows that capture emerging fields. Our TCI is computed on a bipartite network in which corporations are connected to technology classes. A link is present when a corporation is active in a class according to an operational rule, for example filing a sufficient number of patents in that class or meeting an RCA based criterion. When patent counts are sparse, the network becomes thin and the HH iterations that infer complexity from the pattern of connections can become unstable \cite{PintarEssletzbichler2022}.
To mitigate this sparsity problem, future work can refine both the link rule and the information carried by each link. One direction is to move from unweighted to weighted edges that reflect the intensity and impact of inventive activity. Citation based measures are natural candidates, such as raw citations, family normalized citations, or time decayed citations, which may represent technological effort and impact more precisely than simple counts. Another direction is to adopt alternative iterative schemes such as the Fitness Complexity algorithm \cite{Tacchella2012}, which uses non linear iterations to capture structure in sparse settings and may yield more robust rankings \cite{Wu2016,Albeaik2017}.

Building on these methodological directions, future research should examine how such refinements affect the robustness and generality of TCI.
Comparative analyses can evaluate TCI against citation based indicators and alternative weighting schemes.
Machine learning models can then use these enriched features to predict shifts in technological rankings.
Cross national applications will test whether the corporate level stability observed for Japan also appears in other innovation systems.
Ultimately, this study advances the corporate level understanding of technological complexity in Japan.
By identifying high complexity technology fields with greater consistency, it provides actionable evidence for strategic planning and policy design in a period of rapid technological change and intensifying global competition.

%% file: Sections/methods.tex
\makeatletter
\def\@currentlabelname{Methods}%
\makeatother
\label{section:Methods}

\section*{Methods}
\phantomsection
\addcontentsline{toc}{section}{Methods}

\input{Sections/Methods/data.tex}

\input{Sections/Methods/technological_complexity.tex}

%% file: Sections/Methods/data.tex
\subsection*{Data}
\phantomsection
\addcontentsline{toc}{subsection}{Data}
\label{subsection:Data}

The dataset utilized in this study comprises registered patents submitted to the Japan Patent Office (JPO) covering fiscal years 1981 through 2010, encompassing detailed records of 3,189,536 patents filed by 64,622 corporations. Patent records adhere to the International Patent Classification (IPC) system and are further categorized according to the Schmoch classification \cite{Schmoch2008}.

The Schmoch system aggregates IPC codes into 35 distinct technological fields, ensuring balanced class sizes and technological consistency, and has been widely applied in previous research \cite{PintarScherngell2022,Balland2018,Whittle2019}. Nevertheless, the granularity of classification substantially influences analytical outcomes. Administrative classifications vary markedly in granularity and international comparability, potentially introducing analytical noise or biases if inadequately managed \cite{Hidalgo2021}. Employing consistent classification schemes, exemplified by the Schmoch system utilized in this study, mitigates such issues by providing balanced and coherent technological categories for robust analyses.

Traditional methods for measuring patent contributions typically employ raw patent counts \cite{Balland2016,Balland2018,PintarScherngell2022,Pinheiro2022,Whittle2019,Jun2023,Abay2024}. However, such counting methods can lead to overestimation, particularly when patents involve multiple corporations or span multiple technological domains \cite{PintarScherngell2022}. To address potential overestimations, the primary IPC class of each patent is mapped onto the corresponding Schmoch field, with patent value fractionally allocated by equally dividing it among associated corporations and technological fields. This fractional allocation prevents overestimation arising from multi-corporation ownership or cross-domain classifications.

The regional approach evaluates technological complexity as a relative index that reflects regional technological capabilities.
However, this method is sensitive to spatial disparities and tends to yield less stable results when finer classifications such as IPC classes are used, compared to broader classifications like the Schmoch system \cite{PintarEssletzbichler2022}.
Such instability appears in both regional and technological complexity evaluations, as it is inherent in the symmetric structure of the underlying bipartite network.

In this study, we analyze technological complexity at the corporate level to overcome these limitations and achieve a more detailed understanding of regional technological portfolios. 
Additionally, filtering criteria are essential to exclude extreme cases capable of distorting analytical outcomes \cite{Hidalgo2021,PintarScherngell2022}. Accordingly, a filtering criterion based on patent distribution per corporation  is applied in this study, retaining only the top 3\% of corporations in patent output (Fig.~\ref{fig:patentdistribution}). This filtering criterion excludes corporations exhibiting minimal patenting activity, thereby reducing significant analytical noise in network analyses. Following this filtering, the dataset comprises 2,900,359 patents filed by 1,939 corporations, representing approximately 91\% of the original patent volume.

Utilizing these filtered data, technological complexity is based on a bipartite network connecting corporations to technological fields in which they are specialized. Although there are some arguments regarding the optimal specialization index for establishing connections within these networks \cite{PintarScherngell2022}, we follows prior research by employing the Revealed Technological Advantage (RTA) index \cite{Soete1987}.
The RTA index builds upon the revealed comparative advantage framework proposed by Balassa \cite{Balassa1965}. Specifically, the RTA index builds upon the revealed comparative advantage framework proposed by Balassa \cite{Balassa1965}. 
Specifically,the RTA index for a given corporation in a specific technological field measures the ratio between the share of patents held by corporation \(C\) in technology \(T\) and the share of patents in technology \(T\) among all corporations.
The numerator of the RTA index captures the fraction of patent activity devoted to technology \(T\) for corporation \(C\). 
Similarly, the denominator reflects the overall share of patent activity in technology \(T\) among all corporations, computed as the sum of \(W_{CT}\) over all corporations divided by the sum of \(W_{CT}\) over all corporations and technologies.
The RTA index is thus given by

\begin{center} \label{eq1}
    \begin{equation}
    \mbox{RTA}_{CT} = \frac{W_{CT}}{\sum_{T} W_{CT}} \Bigg/ \frac{\sum_{C} W_{CT}}{\sum_{CT} W_{CT}}~.
    \end{equation}
\end{center}
We consider that RTA index values of 1 or greater indicate that corporation \(C\) holds a prominent technological advantage in technology \(T\). Based on this threshold, a binary adjacency matrix of the bipartite network \(M_{CT}\) is defined by

\begin{center}
\begin{equation} \label{eq2}
M_{CT} = 
\begin{cases} 
1 & \text{if } \mbox{RTA}_{CT} \geq 1, \\
0 & \text{otherwise}.
\end{cases}
\end{equation}
\end{center}

%% file: Sections/Methods/technological_complexity.tex
\subsection*{Technological Complexity}
\phantomsection
\addcontentsline{toc}{subsection}{Technological Complexity}
\label{subsection:technologicalcomplexity}
The binary matrix \(M_{CT}\) forms the two metrics which characterize each node in the bipartite network by the sum of connections, known as degree centrality. The diversity of corporations \(K_{C,0}\) quantifies the number of technologies in which a corporation is specialized, and indicated by $M_{CT}$. 
Similarly, the ubiquity of technologies \(K_{T,0}\) represents the number of corporations that exhibit a technological advantage in a given field. These quantities are computed as

\begin{equation}\label{eq3}
\begin{aligned}
K_{C,0} = \sum_{T} M_{CT},  \\
K_{T,0} = \sum_{C} M_{CT}~.
\end{aligned}
\end{equation}

Higher diversity values indicate corporations with broad technological capabilities, while lower ubiquity values signify that the technologies are rare. 
Through calculating these two characteristics of adjacent nodes iteratively, the diversity and ubiquity of each node is updated as

\begin{equation}\label{eq4}
K_{C,N} = \frac{1}{K_{C,0}} \sum_{T} M_{CT}\, K_{T,N-1},
\end{equation}

\begin{equation} \label{eq5}
K_{T,N} = \frac{1}{K_{T,0}} \sum_{C} M_{CT}\, K_{C,N-1}.
\end{equation}

Here, the metric \(K_{T,1}\) represents the average nearest neighbor degree for technology nodes and it reflects the average diversity of corporations connected to a given technology.
Substituting Eq. (\ref{eq4}) into Eq. (\ref{eq5}) yields

\begin{equation} \label{eq6}
K_{T,N} = \sum_{T'} \widetilde{M}_{TT'} K_{T',N-2}
\end{equation}

where \(T'\) denotes a technological field, and $\widetilde{M}_{TT'}$ is defined as

\begin{equation} \label{eq7}
\widetilde{M}_{TT'} = \sum_{C} \frac{M_{CT}\, M_{CT'}}{K_{C,0}\, K_{T,0}}~.
\end{equation}

Eq. (\ref{eq6}) is satisfied when  \(K_{T,N} = K_{T,N-2} = 1\), which corresponds to the eigenvector associated with the largest eigenvalue of the stochastic matrix \(\widetilde{M}_{TT'}\). 
Since this eigenvector is a vector of ones and not informative, the eigenvector corresponding to the second largest eigenvalue of \(\widetilde{M}_{TT'}\) captures the greatest variance among technology fields \cite{Hidalgo2021, Mealy2019}.
Thus, TCI is given by
\begin{equation} \label{eq8}
TCI = \frac{\widetilde{T} - \langle \widetilde{T} \rangle}{\text{stdev}(\widetilde{T})},
\end{equation}

where \(\widetilde{T}\) represents the eigenvector corresponding to the second largest eigenvalue of \(\widetilde{M}_{TT'}\), and \(\langle \widetilde{T} \rangle\) is their mean and \(\text{stdev}(\widetilde{T})\) is their standard deviation.

%% file: Sections/data_availability.tex
\section*{Data Availability}
\phantomsection
\addcontentsline{toc}{section}{Data Availability}
\label{section:data_availability}

The patent data utilized in this study were obtained from the Japan Patent Office (JPO) via the J-platPat portal (\url{https://www.j-platpat.inpit.go.jp/}). Records published within the most recent twelve months are directly downloadable; data older than one year can be provided by the JPO upon reasonable request. Patent classification concordance tables mapping International Patent Classification (IPC) classes to Schmoch technology fields were acquired from the World Intellectual Property Organization (WIPO) IP Statistics portal (\url{https://www.wipo.int/en/web/ip-statistics}).

%% file: Sections/code_availability.tex
\section*{Code Availability}
\phantomsection
\addcontentsline{toc}{section}{Code Availability}
\label{section:code_availability}

All scripts and code used to compute the Technological Complexity Index and reproduce the results of this paper are publicly available in the following GitHub repository: \url{https://github.com/RintaroKARASHIMA/KCIinJapaneseFirms}.  

%% file: Sections/author_contributions.tex
\section*{Author Contributions Statement}
\phantomsection
\addcontentsline{toc}{section}{Author Contributions Statement}
\label{section:authorcontributions}
H.I. conceived the main research idea, designed the study, collected the data, and secured the funding. R.K. performed the analysis and drafted the manuscript. Both H.I. and R.K. critically reviewed the results, discussed the findings, contributed to the revisions of the manuscript, and approved the final version of the manuscript.

%% file: Sections/acknowledgements.tex
\makeatletter
\def\@currentlabelname{Acknowledgements}%
\makeatother
\label{section:Acknowledgements}

\section*{Acknowledgements}
\phantomsection
\addcontentsline{toc}{section}{Acknowledgements}

This work was supported by Japan Society for the Promotion of Science KAKENHI Grant-in-Aid for Scientific Research (B) Grant Numbers JP25K01454 and JP24K00247, and Watanabe Memorial Foundation for the Advancement of New Technology FY2024 Science and Technology Research Grant Number R6-590. 
The funders had no role in study design, data collection and analysis, decision to publish, or preparation of the manuscript.

%% file: Sections/supplementary_information.tex
\section*{Supplementary Information}
\phantomsection
\addcontentsline{toc}{section}{Supplementary Information}

\renewcommand{\thefigure}{S\arabic{figure}}
\setcounter{figure}{0}

\begin{figure}[ht]
    \centering
    \includegraphics[scale=2.75]{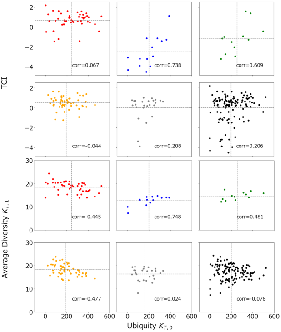}
    \caption{The Pearson correlation between the degree centrality \( K_{T,0} \) of technology node \( T \) classified by IPC and the average nearest neighbor degree \( K_{T,1} \), and the Pearson correlation between the degree centrality \( K_{T,0} \) of technology field node \( T \) and TCI. In both figures, the black lines indicate the mean values. Each data point is color-coded according to the five classifications defined by Schmoch\cite{Schmoch2008}.}
    \label{fig:persector}
\end{figure}

\begin{figure}[ht]
    \centering
    \includegraphics[scale=1.75]{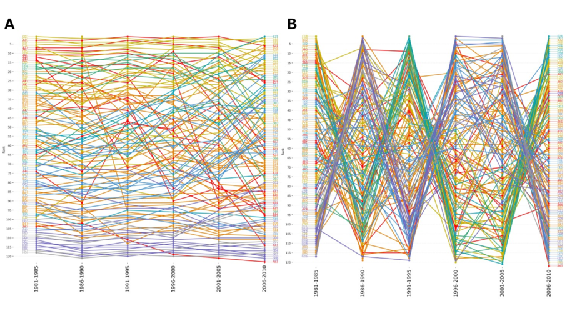}
    \caption{Changes in the five-year rolling ranking of the TCI based on 124 three-digit IPC classes for
      (A) the corporate analysis and
      (B) the regional analysis.
      The lines color-coded according to the one-digit IPC classes.
    }
    \label{fig:bump}
\end{figure}

\begin{figure}[ht]
    \centering
    \includegraphics[scale=0.75]{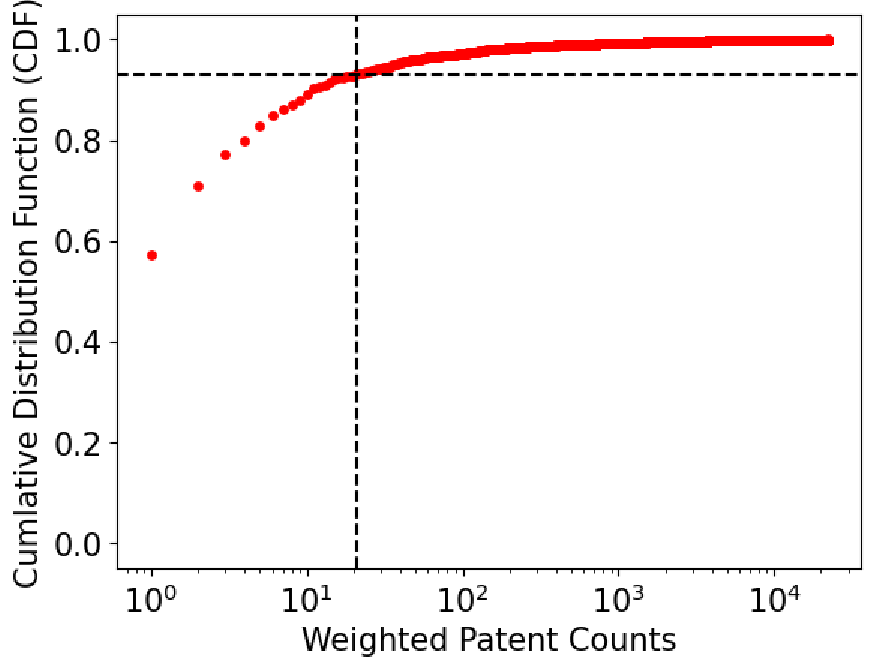}
    \caption{The distribution of weighted patent counts (horizontal axis) for each corporation, with the cumulative distribution function (CDF) of their patents on the vertical axis. The black dashed lines on both axes indicate the excluded region, corresponding to the top 3\% of corporations that account for 93\% of all patents.}
    \label{fig:patentdistribution}
\end{figure}